\documentstyle[12pt,epsfig]{article}
\begin{document}
\title{The role of the quantum properties of gravitational radiation in the detection of gravitational waves.}

\author{ A. Camacho 
\thanks{Email: abel@abaco.uam.mx} \\
Physics Department, \\ 
Universidad Aut\'onoma Metropolitana-Iztapalapa. \\
P. O. Box 55-534, 09340, M\'exico, D.F., M\'exico.}

\date{}
\maketitle

\begin{abstract} 
The role that the quantum properties of a gravitational wave could play in the detection of gravitational radiation is analyzed. 
It is not only corroborated that in the cu\-rrent\- laser--interferometric detectors the resolution of the \-ex\-pe\-ri\-men\-tal\- apparatus could lie very far from the corresponding quantum threshold (thus 
the backreaction effect of the measuring device upon the \-gra\-vi\-ta\-tion\-al\- wave is negligible), 
but it is also suggested that the consideration of the quantum properties of the wave could entail the definition of dispersion of the measurement outputs. 
This dispersion would be a function not only of the sensitivity of the measuring device, but  also of the interaction time (between measuring device and gravitational \-ra\-dia\-tion\-) and of the arm length of the corresponding laser--interferometer. 
It would have a minimum limit, and the introduction of the \-cu\-rrent\- experimental parameters insinuates that the dispersion of the existing 
proposals could lie very far from this minimum, which means that they would show a very large dispersion.
\end{abstract}

\newpage
\section{Introduction.}
The detection of gravitational waves has an important role in contemporary physics 
and the experimental devices that have been employed in this quest have suffered a 
very impressing development. 
Remarkable technological progress has been made in the last few years in order to detect the weak 
gravitational waves that are believed to be emitted from extragalactic sources [1]. 

The measuring process involved in the detection of gravitational radiation using a laser--interferometer entails a continuous interaction 
between the measuring device and a gravitational wave. This feature means that the analysis of this process could be significantly 
simplified with the use of a formalism that allows the explicit introduction of the time evolution of the interaction between measuring device and measured system. 
This is precisely the case of the so called Continuous Quantum Measurements [2].

These experimental efforts are justified by a large amount of comparison \-po\-ssi\-bi\-li\-ties\- that will emerge if these waves 
could be measured. For instance, by comparing the arrival time of the first bursts of light and 
gravitational waves from a distant supernova, one could verify \-ge\-ne\-ral\- relativity's 
prediction that electromagnetic and gravitational waves propagate with the same speed, with 
other words, that they couple to static gravity (spacetime structure) of our Galaxy and other 
galaxies in the same way. By measuring the polarization properties of gravitational waves, one could verify general 
\-re\-la\-ti\-vi\-ty's\- prediction that these waves are tranverse and traceless, and thus are the classical 
consequences of spin--two gravitons [3]. The black hole no hair theorem could also be tested, as well as for searches for 
exotic stellar objects, such as boson stars, soliton stars, and naked singularities [4]. 
By comparing the detailed wave forms of observed 
gravitational radiation bursts with those predicted for the coalescence of black hole binaries, one could also verify that certain bursts 
are indeed produced by black hole coalescence, and as a consequence, verify unequivocally the 
existence of black holes. Such verifications would constitute by far the strongest test ever of Einstein's 
laws of gravity.

Nevertheless, the interest in gravitational waves does not finish with the \-po\-ssi\-bi\-li\-ty\- that they offer to compare 
general relativity's predictions with the experiment. They also enable us to confrontate with the experiment the cosmological model 
of inflation, which postulates a period of accelerated expansion in the universe distant past [5]. Although 
originally contemplated to give an explanation to a host of cosmological conundrums such as 
the horizon, flatness and monopole problems, one of its more useful properties is that it generates a spectra 
of gravitational waves [6], which originate anisotropies in the microwave background [7]. These anisotropies have already been 
measured by the 
COsmic Background Explorer (COBE), and the new \-ge\-ne\-ra\-tion\- of satellites like PLANCK (former COBRAS/SAMBA) [8], which will have a higher sensitivity 
and resolution than COBE, will be able to measure these anisotropies with a much higher precision. 
Hence, they will offer a key test of the \-in\-fla\-tio\-na\-ry\- picture, because they will render a very precise 
picture of these \-tem\-pe\-ra\-tu\-re\- anisotropies induced by gravitational waves. 
But that is not the whole story, the detection of gravitational waves offer the opportunity not only to check the 
\-in\-fla\-tio\-na\-ry\- prediction but also to rule out some inflationary models [9].  It might also help to reconstruct 
the global topology of the universe, by means of the analysis of the effects of these waves upon the cosmic microwave background [10].

The experiments in this research can be classified in two major categories: those using laser--interferometers with very long arms, 
and those using resonant bars cooled to ultra--low temperatures [11]. There are several problems that appear in these kind of projects. 
For instance, several noise sources emerge in these experimental 
devices, thermal motion of the bar constituents, electronic noise, seismic noise, shot noise (due to photon--counting 
fluctuations in a photodetector) [1]. But there is an additional element that complicates these experiments, the so called quantum limit, where this means, 
for instance, 
the quantum limit in the sensitivity of the beam receiver of a laser--interferometer, this sensitivity has its ultimate source in the vacuum fluctuations 
of quantum electrodynamics [12]. 

It is noteworthy to comment that the aforementioned quantum properties are \-a\-sso\-cia\-ted\- to the measuring device, and at this point we may wonder, if the quantum properties of a gravitational 
wave could play an important role in the emergence of some kind of problem in this detection process. The common argument reads: these waves behave exceedingly classically; quantum 
mechanical corrections to the classical theory have fractional magnitude $\sim 10^{-37}$ [1]. Nevertheless, 
there may exist \-ex\-pe\-ri\-men\-tal\- situations in which the quantum behaviour of the gravitational field could not be so easily
neglected, for instance, in the case of the measurement of length, \-gra\-vi\-ta\-tio\-nal\- fluctuations may dominate those who 
have as origin the measuring device, if macroscopic masses, i. e. masses larger than the Planck mass, are used [13]. In other words, 
we may have some measurement situations in which the noise coming from the measuring device is smaller than the noise 
that has its origin in the quantum properties of the gravitational field. This last result suggests that the exclusion of the quantum properties of the measured system
from the analysis of the measurement process of gravitational radiation could be sometimes not soundly justified, and that its inclusion could render a more realistic picture of this process.
 
We will contemplate the detection process of gravitational radiation taking into account its quantum properties, 
namely we consider this process from the point of view of the so called continuous quantum 
measurements [2]. As has already been \-co\-mmen\-ted\-, this formalism contains as one of its features the explicit time behaviour of the 
interaction between measuring device and measured system. 
The quantum \-cha\-rac\-ter\- of the measured system and the aforementioned time evolution are included in the mathematical treatment of the problem 
through the use of the so called Restricted Path Integral Formalism (RPI) [14], which considers the interaction between 
measuring device and measured system in the form of a weight functional and introduce it in Feynmann's 
path--integral formalism of the measured system. 

The analysis of the measurability of the electromagnetic field [15] in the context of RPI renders some interesting  
results, 
for example, the characteristic time and length of the experiment play an important role in the definition of the 
dispersion of 
the measurement outputs. If we remember that a gravitational wave can be seen as a tensor, which satisfies the wave 
equation and has the same 
propagation speed as light, defined on a Minkowskian spacetime, then we could expect to obtain in the analysis of the measurability 
of gravitational radiation something similar to the results of the electromagnetic case. 

The analysis of all the possible noise sources 
is a very important task, because the interferometers now under construction will operate with a sensitivity that does not 
guarantee that they will see any sources, and almost certainly what they will see will be difficult to dig out of the noise [16]. 
If we at this point remember Jaekel and Reynaud's work [13], which points out the importance in some measuring processes of the quantum properties of the gravitational field 
in the emergence of fluctuations (which can be larger than those that have as origin the measuring apparatus), then we may wonder if the noise source that emerges when we take into account 
the quantum properties of a gravitational wave could not, in some cases, become so strong, that they should not be neglected in the context of the measurement process of gravitational radiation.

In the present work we will analyze the measurability of a monocromatic plane gravitational wave by means of RPI, namely we will 
take into account the quantum behaviour of gravitational radiation. 
The employed formalism allows us to consider the effects of a measurement process without introducing a particular scheme for 
the measuring device, in other words, it is a phenomenological approach. 

We will show not only that a gravitational wave behaves, indeed, exceedingly \-cla\-ssi\-ca\-lly\-, but also suggest that 
a minimum limit could exist in the dispersion of the measurement outputs, the one could not be overcome. This limit in 
the dispersion would be a function of the four--dimensional volume in which the experiment is carried out. Thus, the 
characteristic time and charateristic length of the experiment could be considered as unrelated parameters, 
but the prize to be paid would be an experiment in which its precision would not be the best one, i. e. the measurement outputs might have a large dispersion. 
This behaviour of the dispersion in the measurement outputs has its origin in the quantum properties of gravitational 
radiation. 
The understanding of this point becomes relevant because in the current stage of the development of laser--interferometers, one of the most important modifications that should render better experiments comprises an increase (with respect to the present typical measuring times) of the interaction time between measuring device and gravitational radiation.

We will find three measurement regions, namely a classical region, a quantum region and a region that lies between the first two, which also defines the minimum of the dispersion. 
This minimum entails the definition of a parameter that could allow us to obtain the relation between the experimental features of our 
measuring device that could render the smallest dispersion. 

With this new parameter we may find: (i) a quantitative analysis of the dispersion of the measurement outputs of any experimental apparatus, (ii) compare the dispersion from different experimental proposals and find out which one of them has the smallest one, (iii) it also enables us comprehend if the introduction of some change or changes in the current experimental proposals leading to a reduction of their dispersions could be feasible under the present technological limitations.

The results are applied on laser-interferometric detectors and we will find; (i) the dispersion associated to them, (ii) the relation between the experimental parameters that could lead us to the smallest physically possible dispersion of the measurements outputs, (iii) we will also be able to see that nowadays any change in this type of proposal leading to a reduction of the dispersion could not be feasible. 

This papers is organized as follows. In Section 2 we formulate in terms of RPI the dynamics of a gravitational wave under the action of a continuous measurement process. 
We introduce in section 3 a definition of dispersion for a gravitational waves detector that stems from our approach, and apply it upon one of the currently most 
discussed detectors, a laser--interferometer. Section 4 contains the conclusions and final discussions of our work.
\bigskip
\section{Formulation of the problem.}
\bigskip

The starting point is the action function of a field $h$

\begin{equation}
S[h] = \int_{\Omega}d^4xL(h, \partial h).
\end{equation}

The dynamics of this field is given by 

\begin{equation}
 A = \int d[h]exp(iS[h]),
\end{equation}

\noindent where the integral has to be performed over all the field configurations. It is important to mention at this point that we are employing Planckian units, i. e. $ G = 1$, $c = 1$, and $\hbar = 1$.

The dynamics of the field is given by this last expression only when no measuring process is present, clearly all the configurations then have the same \-pro\-ba\-bi\-li\-ty\-.
The presence of a measuring device and its use leads to the break down of the \-e\-qui\-pro\-ba\-bi\-li\-ty\- of the configurations of the field. 

We may express the influence of the measuring device on the dynamics of the field by restricting the path integral [2]. The information corresponding to a given experimental output $a$ may be expressed by a weight functional $\omega_a[h]$. 
This weight functional introduces the condition of a continuous measurement [14].

\begin{equation}
A_a = \int d[h]\omega_a[h]exp(iS[h]).
\end{equation}

The more probable the configuration $[h]$ is, according to the output $a$, the bigger that $\omega_a[h]$ becomes [15], 
this means that the value of $\omega_a[h]$ is approximately one for all field configurations $[h]$ that agree with the measurement output $a$ and it is almost 0 for those configurations that do not match with the results of the experiment. 

Expression (3) may be employed to find the probabilities associated with \-di\-ffe\-rent\- measurement outputs. We need only to calculate this expression under the same boundary conditions and same integration region but taking into account different measurement results $a$.
The aforementioned probability density for the different \-po\-ssi\-ble\- measurements outputs is then $P_a = \vert A_a \vert^2$. 

We will take a metric with the form $g_{\mu \nu} = \eta_{\mu \nu} + h_{\mu \nu} $, under the following condition: $\vert h_{\mu \nu} \vert <<1$, and evaluate our physical quantities up to second order in $h_{\mu \nu}$. 
In our case the Lagrangian is $L[h] = \sqrt{-g}R$, where $R$ is the scalar curvature and $g$ the determinant of the metric [17].

The scalar curvature is then given by

{\setlength\arraycolsep{2pt}\begin{eqnarray}
R \approx \partial^{\mu}\partial^{\nu}h_{\mu \nu} - {1\over 2}\partial^{\mu}\partial_{\mu}h + {1\over 4}\partial^{\mu}\partial_{\mu}(h^{\tau \nu}h_{\tau \nu}) - {1\over 2}\partial^{\mu}\partial_{\tau}(h^{\tau \nu}h_{\nu \mu}) + {1\over 2}\partial^{\mu}(h_{\mu \nu}\partial_{\tau}h^{\tau \nu}) \cr 
{1\over 2}h^{\tau \nu}\partial_{\tau}\partial_{\nu}h + {1\over 2}(\partial_{\nu}h^{\tau \nu})(\partial_{\tau}h) - {1\over 4}(\partial_{\tau}h)(\partial^{\tau}h) + {1\over 2}(\partial_{\tau}h)(\partial_{\nu}h^{\tau \nu}) - (\partial^{\mu}h_{\mu \tau})(\partial_{\nu}h^{\tau \nu}) \cr 
+ {1\over 4}(\partial^{\mu}h_{\nu \tau})(\partial_{\mu}h^{\nu \tau}), 
\end{eqnarray}

\noindent where $h = h^{\nu}~_{\nu}$.

We know that in the linearized theory of general relativity the components of the Riemann tensor, and in consequence also the scalar curvature, are invariant under gauge transformations.
Therefore, in order to simplify (4) we proceed now to define the fields $\bar{h}_{\mu \nu} = h_{\mu \nu} - {1\over 2}\eta_{\mu \nu}h$. If we now introduce the following  four ``gauge conditions'' $\partial^{\nu}\bar{h}_{\mu \nu} = 0$, which are sometimes denoted as the Lorentz Gauge, then the curvature becomes 

\begin{equation} {R \approx - {1\over 2}(\partial^{\mu}\bar{h}_{\nu\tau})(\partial_{\mu}\bar{h}^{\nu\tau}) }. \end{equation} 

The action is simplified and it takes now the form

\begin{equation} { \int_{\Omega}Ld^4x = - {1\over 32\pi}\int_{\Omega}d^4x(\partial^{\mu}\bar{h}_{\nu\tau})(\partial_{\mu}\bar{h}^{\nu\tau}) }. \end{equation} 

The amplitude becomes 

\begin{equation} { A = \int d[\bar{h}]\delta(\partial^{\nu}\bar{h}_{\lambda \nu})exp[-{i\over 32\pi}\int_{\Omega}d^4x(\partial^{\mu}\bar{h}_{\nu\tau})(\partial_{\mu}\bar{h}^{\nu\tau})] }. \end{equation} 

The integration must be performed over those configurations that comply with Lorentz Gauge. The fulfillment of this condition in 
expression (7) is guaranteed with the introduction of Dirac's delta. 
The integral appearing in the last expression has to be carried out in the four--dimensional volume in which the gravitational wave is 
located, we denote it by $\Omega$. 

In this work we will always assume that the condition that allows the definition of a gravitational wave is fulfilled, 
the so called ``short wave approximation'' [18]. This restriction means that the lengthscale on which the waves vary is very short compared to the lengthscales on 
which all other important curvatures vary, and in astrophysical situations this is a very realistic assumption [19].

Let us define the following vector field: ${\bf E}_{\bar{h}}^2 = -{1\over 8\pi}(\partial^{\mu}\bar{h}_{\nu\tau})(\partial_{\mu}\bar{h}^{\nu\tau})$. 
In term of this vector field our amplitude can be rewritten as 

\begin{equation} { A = \int d[\bar{h}]\delta(\partial^{\nu}\bar{h}_{\lambda \nu})exp[{i\over 4}\int_{\Omega}{\bf E}_{\bar{h}}^2d^4x] } .\end{equation} 

Expression (8) implies a path--integral quantization of General Relativity. This idea was developed by Ha\-lli\-well\- [20] and it was also 
proved that it is equivalent to Wheeler--DeWitt equation, which results from a quantization scheme {\em \'a la} Dirac.  
At this point we must add that our last expression is valid only when we do not perform an experiment in the \-con\-si\-der\-ed\- region $\Omega$ [2, 15]. 

Let us now suppose that we measure, continuously in time, the field ${\bf E}_h(x)$ and that this measurement process renders as measurement output the field ${\bf E}(x)$.

We introduce at this point an additional hypothesis; the interaction between measuring device and gravitational wave is characterized by the following functional

\begin{equation} 
{ \omega_{[{\bf E}]}[\bar{h}] = exp\{ -{1\over \tilde\Omega}\int_{\tilde\Omega}{({\bf E}_{\bar{h}} - {\bf E})^2\over \Delta E^2}d^4x\},   } 
\end{equation} 
\bigskip

\noindent where $\tilde\Omega$ is the volume of the four--dimensional region in which the experiment is going to be carried out and it is not the integration volume that appeared previously, which we denoted by $\Omega$.

The interaction that arises between wave and measuring device is restricted to the region of spacetime in which the measurement process takes place, namely to the four--dimensional volume $\tilde\Omega$. 
Therefore, with the presence of a measurement process the four--dimensional volume in which the integration has to be performed becomes now the four--dimensional volume in which the measuring interaction occurs, namely $\tilde\Omega$ [2, 15].

The concrete form of the weight functional $\omega_{[{\bf E}]}$ depends on the measuring device [21], and clearly we do not know if the involved experimental constructions have these type of weight functionals.
But our goal is two-fold, we look not only for an estimation of the measurability of the gravitational waves but also try to take into account the effect of the quantum properties of the gravitational radiation upon the measurement process. In a first approach to these two issues, we may consider (9) as an approximation to the actual weight functional. 
The back influence of the measuring device on the field is also already taken into account in (9).
The form given by (9) has another very attractive feature, it leads us to gaussian integrals, and these type of integrals can be easily calculated.  

 The integration here must be done in $\tilde\Omega$, which is the four--dimensional volume where the interaction between wave and measuring device takes place. 
Therefore, ``characteristic time'' means the time in which the measuring device interacts with the wave, namely it is the time lapse the experiment lasts. 
The phrase ``characteristic length'' denotes in the case of a laser-interferometer the arm length of the apparatus, for example in the case of the {\it Virgo} project [22] it is $l = 3km$.

The quantity $\Delta E^2$ may be intepreted as the resolution of the experimental device [15], associated to the field {\bf E}.

We may find from the following expression the probability associated to a measurement process which renders as output the field configuration ${\bf E}$, $P_{[{\bf E}]} = |A_{[{\bf E}]}|$, where
 
\begin{equation} { A_{[{\bf E}]} = \int d[\bar{h}]\delta(\partial^{\nu}\bar{h}_{\lambda \nu})exp\{{i\over 4}\int_{\tilde\Omega}{\bf E}_{\bar{h}}^2 d^4x - {1\over \tilde\Omega}\int_{\tilde\Omega}{({\bf E}_{\bar{h}} - {\bf E})^2\over \Delta E^2}d^4x\} }. \end{equation} 

This last functional integral can be easily calculated employing the usual trick of shifting to zero boundary conditions [15]. 
Let $h_{\mu \nu}^{(c)}$ be an extremal (classical) field configuration, satisfying a given boundary condition at $\partial\Omega$ 
and let ${\bf E}_c$ be the \-co\-rres\-pon\-ding\- strength field. In other words, here $h_{\mu \nu}^{(c)}$ is the field configuration that minimizes (6), and therefore is 
a solution of the linearized Einstein equations. 

This definition enables us to write the fields appearing in any possible measurement output as the sum of two contributions; (i) the first term is the classical field, it comes from the minimization of expression (6) and corresponds to the solution of the Einstein equations, and (ii) the second term is a deviation with respect to the classical result. 
Hence, we may write, ${\bf E}_h = {\bf E}_c + {\bf E}_f$.

The importance of this definition lies on the fact that, as we will see later, the \-pro\-ba\-bi\-li\-ty\- of finding in the 
measurement outputs a certain value of {\bf E} will be a function of the norm of the deviation of the corresponding field with 
respect to the classical field. 

Taking into account these definitions (10) becomes

\begin{equation}
 A_{[{\bf e}]} = exp\{{i\over 4}S[h^c]\}\int d[f]\delta(\partial^{\nu}f_{\mu \nu})exp\{\int_{\tilde\Omega}(i{\bf E}^2_f - {1\over\tilde\Omega}{({\bf E}_f - {\bf e})^2\over \Delta E^2})d^4x \} .
\end{equation}
Here ${\bf e}$ is the configuration of ${\bf E}$ but reduced to the null zero boundary condition, namely ${\bf E} =
 {\bf E}_c + {\bf e}$.

Let us now introduce the definition $\alpha = {\tilde\Omega \Delta E^2\over 4}$, ${\bf E}_t = {\bf E}_f + {1\over 1 - i\alpha}{\bf e}$.  
From this new variable

{\setlength\arraycolsep{2pt}\begin{eqnarray}
A_{[{\bf e}]} = exp\{{i\over 4}S[\bar{h}^c]\}exp\{{i\over 4}\int_{\tilde\Omega}({1\over 1 - i\alpha}{\bf e}^2 d^4x\} 
\times\nonumber\\
\int d[f]\delta(\partial^{\nu}f_{\mu \nu}) exp\{ {i\over 4}\int_{\tilde\Omega}[(1 + {i\over \alpha}){\bf E}^2_t \}.
\end{eqnarray}} 
\bigskip

Let us now change the integration variable $f_{\mu\nu}$ to $t_{\mu\nu}$, the latter being the potential corresponding 
to ${\bf E}_t$. It is readily seen that the functional integral does not depend on the fields ${\bf e}$. Hence, it will appear as a normalization constant, the one does not interest us.

Therefore, the probabilty density, up to a normalization constant, associated to the configuration ${\bf E}$ is 

\begin{equation} { P_{[{\bf e}]} = \vert A_{[{\bf e}]}\vert^2 = exp\{-{1\over 2}\int_{\tilde\Omega}({1\over \alpha + \alpha^{-1}}{\bf e}^2 d^4x\}  }. \end{equation} 

This last expression may be reformulated as follows

\begin{equation} { P_{[{\bf E}]} = exp\{-{2\over \tilde\Omega}\int_{\tilde\Omega}{({\bf E}_c - {\bf E})^2\over {16\over\tilde\Omega^2 \Delta E^2} + \Delta E^2}d^4x\}  } .\end{equation} 

Expression (14) tells us that the measurement of ${\bf E}(x)$ most probably results in a configuration which coincides with 
the classical one, namely with ${\bf E}_c(x)$. 

Nevertheless, the results may differ from the classical case and at the same time have a probability close to one, 
that is precisely the situation if the probability density remains close to its maximum value, in other words, if we have that

\begin{equation} {{1\over \tilde\Omega}\int_{\tilde\Omega}[({\bf E}_c - {\bf E})^2]d^4x < \delta E^2 } ,\end{equation} 

\noindent where we have the following definition

\begin{equation} { \delta E^2 = {16\over\tilde\Omega^2 \Delta E^2} + \Delta E^2 } .\end{equation} 

\bigskip
\section{Application to gravitational waves detector.}
\bigskip

\subsection{Dispersion of a gravitational waves detector.}

Expressions (15) and (16) define the dispersion of the measurement outputs. 
Let us explain this a little bit better. If we could perform these type of experiments several times and plot the values of $ {\bf E}^2$ against the number of times that each one of these values was obtained, then we would obtain a graph that could be seen as a probability graph for $ {\bf E}^2$. Here the word dispersion means the variance of this probability curve.

Expressions (15) and (16) also enable us to define three different regions of measurement:

(a) The first one corresponds to the case $\Delta E^2 > 4/\tilde\Omega$, hence, $\delta E = \Delta E$. 

The dispersion of the measurement outputs has its origin only in the resolution of the experimental device. That means, we perform a classical mode of measurement. The dispersion decreases if the measuring device becomes more precise. 

(b) The second region is defined by $\Delta E^2 < 4/\tilde\Omega$. In consequence, $\delta E = 4/(\tilde\Omega\Delta E)$.

It is easy to see, that with the decrease of $\Delta E$ the dispersion increases. We are now in a quantum mode of measurement. 
With other words, improving the experimental resolution means that the backreaction effect of the measuring device upon the field becomes more important and the information we receive contains more influence of the experimental apparatus.
 
This is an unavoidable effect, and it imposes a maximum limit to the objective information that from the wave can be extracted. Clearly Heisenberg's Uncertainty Relation stands behind this effect.

(c) The third region, which lies between the classical and quantum modes of measurement, corresponds to the case $\Delta E^2 = 4/\tilde\Omega$. 
It is readily seen that this case yields a minimum in the dispersion of the measurement outputs. Therefore, we might use 
this last result as a criteria and with it find the relation between the experimental parameters that could render the smallest variance in our probability curve for ${\bf E}^2$, i.e. the smallest dispersion in the measurement outputs.

 Let us denote the characteristic time and length of the experiment by $\tau$ and $l$, respectively. Hence, $4/\tilde\Omega = 4/(l^3 \tau)$, and the minimum in the dispersion emerges if we have $\Delta E^2 = 4/(l^3 \tau)$. 

Here we obtain a criteria that could allow us to evaluate in a quantitative way the dispersion or variance of the measurement outputs. 
A small dispersion would mean, therefore, $\Delta E^2l^3\tau/4 \sim 1$, while $\Delta E^2l^3\tau/4  < 1$ would be associated to the case of a quantum measurement, and $\Delta E^2l^3\tau/4 > 1$ to the classical measurement regime.
It also enables us, as we will later see, to determine if we might introduce some change or changes in these type of experiments which could lead to a reduction of the dispersion. 

Our result allows us also to compare two different experimental constructions and calculate which one of them will show the smallest dispersion. 
Let us explain this a little bit better. If we have two different experimental proposals, such that their parameters satisfy $\Delta E_1^2l_1^3\tau_1/4 > \Delta E_2^2l_2^3\tau_2/4 > 1$, then in the first experiment the criteria that evaluates the dispersion is farther from 1 than the criteria of the second experiment, and in consequence the first one has also a larger dispersion. 

Our result insinuates also that the characteristic time and length of the experiment do indeed have an important contribution to its dispersion and that increasing the time, while keeping the length constant, reduces the quantum threshold and in consequence makes an experiment ruder, in other words, it becomes a worse experiment, because the dispersion of the measurement outputs grows. 

We may understand this last result better with an example. Let us suppose that with the {\it Virgo} project [22] we perform two different sets of experiments. In the first one we let each one of the experiments run a typical time [11] $\tau = 100h \sim 10^{5}s$. In the second one the experiments last $\tau = 1000h \sim 10^{6}s$. 
Then, our result says that in the first set the variance of the obtained probability curve for ${\bf E}^2(x)$ would be smaller than the dispersion of the second set. Indeed, in both experiments the experimental resolution and charateristic length are the same, but  we have $\tau_1 < \tau_2 \Rightarrow \Delta E^2l^3\tau_1/4 < \Delta E^2l^3\tau_2/4$ (we will show below that both measurement processes are classical ones,  this assertion means that $1 < \Delta E^2l^3\tau_1/4$), therefore, the experiment that lasts $\tau_1$ has a smaller dispersion than the one that has as characteristic time $\tau_2$. 

To resume, the characteristic time and characteristic length of the experiment could be not independent from each other, 
at least in the context of the detection of gravitational radiation. 
We might play with them but the result could be that the dispersion of the corresponding measurement outputs would lie very far from its minimum.

\subsection{Application to a laser--interferometer.} 

Let us now apply these results to the case of a laser-interferometric detector. 
Translating our expression to ordinary units we have that the minimum appears if we have $\Delta E^2 = 4l_p^2/(l^3 c\tau)$, where $l_p = \sqrt{G\hbar /c^3}$ is Planck length.

We have written the scalar curvature as $R \approx {\bf E}^2(x)$ (see eqno. (5) and the \-de\-fi\-ni\-tion\- that appears before eqno. (8)). Hence, the quantity $\Delta E^2$ is related to the resolution of the measuring device with respect to the scalar curvature.

Let us now consider the average scalar curvature $R_a$ of a gravitational wave that has a frequency $\nu_g$ and an intensity $I$, 
here average means that we integrate over one period. Then we have that $R_a ~\sim I\nu_g^2/c^2$ [17]. 

One of the most important limitations that emerge in connection with the use of laser--interferometric detectors is related to their sensitivity. 
The current stage of this parameter enables us to measure intensities at most of $I\sim 10^{-21}$ [11, 23], 
in other words, their resolution in the intensity is, in the best case, equal to $\Delta I \sim 10^{-21}$. This last fact implies that the resolution in the average scalar curvature 
is $\Delta R \sim \nu_g^2\Delta I/c^2$.

From our results we have that the minimum in the dispersion would appear if $4l_p^2/(l^3c\tau) \sim \nu_g^2 \Delta I/c^2$.
Hence, we may define the dispersion in these experiments as \-fo\-llow\-s\-

\begin{equation}
\delta R = 4{cl_p^2\over l^3\tau\nu_g^2\Delta I}.
\end{equation}
\bigskip

A small dispersion means $\delta R \sim 1$, while an experiment falls in the classical region if $\delta R < 1$, and in the quantum region 
when $\delta R > 1$. It is noteworthy to mention that the dependence of the dispersion in the experimental parameters does not consist 
only of the sensitivity $\Delta I$, the quantum properties of the gravitational wave introduce in this definition the characteristic time and length of the experiment, 
in other words, $\tau$ and $l$ appear in expression (17) because we have considered the quantum properties of the wave. 
This definition suggests also how to relate the experimental parameters in order to minimize the dispersion of the measurement outputs. 
Having an estimation of the intensity of the gravitational wave, of the characteristic length of the measuring device, 
and of the resolution in the intensity (these last two parameters are determined by the physical construction of the experimental apparatus) 
we may calculate the relation between the time of the measurement process and the wave frequency, at which we should aim, 
in order to reduce the dispersion. 
In other words, according to our results, the experiment with the smallest dispersion would require the fulfillment of the following 
condition; $\tau\nu_g^2 \sim 4cl_p^2/(l^3\Delta I)$.

Let us now consider the case of the {\it Virgo} project [22] in which $\nu_g = 100$Hz, $\Delta I = 10^{-21}$, and $l = 3$km. 
Considering $I = 10^{-18}$ (the possible intensity if a neutron star of our galaxy suffers a gravitational colapse, with an energy emission in form of gravitational radiation 
equal to $0.1M_{\odot}c^2$, being $M_{\odot}$ the solar mass [23]) we obtain the following condition 
$4cl_p^2/(l^3\Delta I) \sim 10^{-53}(seconds)^{-1}$. 
On the other hand, if we wish a small dispersion of the measurement outputs, then the characteristic time of this experiment 
should be $\tau \sim 10^{-53}s^{-1}/(100Hz)^2 = 10^{-57}$ seconds. This lapse time is thirteen orders of magnitude smaller than Planck time !! 

If we now try to modify the length of these proposals, and attempt to obtain a dispersion close to the minimum, then we have 
that $l^3 = 4cl_p^2/(\nu_g^2 \tau\Delta I)$. Employing $\tau = 100h \sim 10^{5}s$, $\nu_g = 100$Hz, and $\Delta I = 10^{-21}$, we find that $l \sim 10^{-14}$cm, a length that is 5 orders of magnitude smaller that Bohr radius, obviously 
this experiment could not be carried out. 

The above estimations suggest not only that any practical realization of this experiment would take place in the classical region but also that the 
dispersion in the measurement outputs could be very large. 
For instance, if we consider a typical measuring time [11] $\tau = 100h \sim 10^{5}s$, 
then, introducing it in our expressions, we obtain that $\Delta R = \nu_g^2 \Delta I/c^2 \sim 10^{-35}$(cm)$^{-2} >> 4l_p^2/(l^3 c\tau) \sim 10^{-96}$(cm)$^{-2}$, we are indeed in the classical region. 
In other words, concerning the magnitude of the back\-re\-ac\-tion\- e\-ffect\- upon the gravitational radiation, the present results show that these experiments 
behave, as expected, exceedingly classically, i. e. the resolution of the measuring device lies very far from the quantum threshold, 
and in consequence the backreaction \-e\-ffect\- upon the gravitational wave may be neglected. 
But this work also insinuates that the role that the quantum properties of the measured system could play in the measuring process of gravitational radiation could not 
be restricted to the ratio between resolution of the experimental apparatus and quantum threshold. The introduction of the nonclassical properties of the wave 
would allow us to associated a dispersion to the measurement outputs, and the introduction of the current \-pa\-ra\-me\-ters\- related to the case of a laser--interferometers 
suggests that they could have a very large dispersion, $\delta R \sim10^{-61}$ (see eqno. (17)).

\bigskip
\section{Conclusions.}
\bigskip

Our system has been a gravitational wave. Taking into account only up to second order perturbations in the metric we have calculated the Lagrangian of this system. 

We have considered the quantum properties of gravitational radiation in the measuring process of a gravitational wave. The effect of the measuring device upon the wave has been 
introduced by means of the Restricted Path Integral Formalism. We have been able to calculate the probability density associated to a continuous measurement in time of the perturbation of the metric, 
up to a normalization constant, 
where this probability density is related to the possibility of having as a measurement output a given configuration of the perturbation of the metric. 
The maximum of this probability density is located around the ``classical'' result, namely the field \-con\-fi\-gu\-ra\-tion\- that is solution of the linearized Einstein equations. We obtained also an expression for the dispersion of the measurement process, the one allows us to define three different regions of measurement:

a) The classical region, in which the dispersion has its origin only in the resolution of the experimental device. A more precise measuring device renders a smaller dispersion.

b) A quantum region that imposes an unavoidable limit to the minimum of the dispersion in the measurement outputs.   
Improving the experimental resolution of the measuring device originates a larger dispersion that emerges as consequence of the backreaction effect of our measuring device upon the gravitational wave. With other words, it sets a limit to the objective information that we may extract from the gravitational wave.
 
c) The third region lies between the two aforementioned ones and determines the \-mi\-ni\-mum\- of the dispersion. This minimum is a function of the four--dimensional \-vo\-lu\-me\- of the region in which the experiment is carried out 
and displays the intrinsic nonseparability of the characteristic time and length of the experiment in connection with the dispersion of the measurement outputs.
 This result enables the definition of a criteria that could allow us to find the relation between the experimental parameters of the measuring device that would render the smallest possible dispersion of the measurement outputs. 

Employing this parameter we may find; (i) the dispersion associated to any of these experimental proposals, (ii) try to see if we may introduce some change in the current proposal that could lead to a reduction of its dispersion, and (iii) compare the dispersion among the different proposals that in this direction already exist and see which one of them has a dispersion closer to its minimum. 

As an application of these results we considered the specific and very important case of a \-la\-ser--in\-ter\-fe\-ro\-me\-tric detector and our results suggest that, for any practical \-rea\-li\-za\-tion\- of these experiments, the measuring process would take place in the \-cla\-ssi\-cal\- region. 
But the present approach allows us to analyze another point, namely we may give a quantitative e\-va\-lua\-tion\- of the effect of the quantum properties of the wave upon the dispersion of the measurement outputs. 
According to our results a small dispersion fulfills the already known relation $\Delta R \sim 4l_p^2/(l^3 c\tau)$, and in the proposed experiments these two parameters would satisfy the condition $\Delta R \sim [4l_p^2/(l^3 c\tau)]10^{61}$. 
Therefore, these\- ex\-pe\-ri\-ments\- could have a very large dispersion, i. e. expression  (17) means that $\delta R \sim 10^{-61}$. 
The explanation to this last result lies on the fact that the the product $1/(l^3\tau\Delta I)$ is very small, $\sim 10(cm^3s)^{-1}$, a 
dispersion close to its minimum needs $1/(l^3\tau\Delta I) \sim 10^{60}(cm^3s)^{-1}$.

The possibility of a dispersion close to the minimum, in connection with the use of laser--interferometric detectors, imposes the condition of a 
\-cha\-rac\-te\-ris\-tic\- observation time that has to be much smaller than the Planck time, a senseless condition, and a modification in the arm length that could lead us to a significant reduction of the dispersion would be also impossible. 

In relation with the role that the characteristic measuring time plays in the \-de\-fi\-ni\-tion\- of the dispersion of the measurement outputs 
it is noteworthy to comment, that in laser--interferometric detectors a phase delay between two light fields having propagated up and down two perpendicular directions 
is actually measured. The measured phase difference can be increased by increasing the arm length $l$, or \-e\-qui\-va\-len\-tly\-, 
the interaction time $\tau$, 
of the measuring device with the gravitational wave. This need appears because the phase delay $\Delta \phi$, has essentially the form 
$\Delta \phi = 4\pi nlI/\lambda_c$, where $n$ is the number of round--trips of the laser beam, and $\lambda_c$ is the wavelength of 
the employed light [1]. These proposals try to compensate the smallness of $I$  with a large $nl$, which, by the way, is equal to $c\tau$. 
It is very unpractical to build large \-in\-ter\-fe\-ro\-me\-ters\-, therefore the solution seems to be an increase in the interaction time [24]. 
On one hand, we know that the resolution $\Delta R$ of these devices is a fixed parameter, but on the other one, 
we have that the minimum dispersion 
would emerge if the condition $\delta R \sim 1$ is \-sa\-tis\-fied\-. From the previous results we suspect that the current 
experiments could not fulfill it. 
If we now increase the interaction time (while the remaining parameters do not suffer any kind of change), 
namely we modify only  $\tau$, then $\delta R$ (which is already much smaller than 1) would become smaller (see eqno. (17)), 
which means that 
with an increase of the interaction time, the dispersion of the measurement outputs of the current experimental devices would
grow. 
Moreover, in order to obtain in these type of experiments a detectable $\Delta \phi$, the interaction time between measuring device and gravitational wave has to become larger, but this condition would render as a \-by\-pro\-duct\- a larger dispersion of the measurement outputs. 
This last conclusion suggests, that even though that the resolution of these experimental devices could lie very far from their quantum threshold, 
which means that the in the context of the detection of gravitational radiation the backreaction effect upon the gravitational wave would be very small, 
the role that nonclassical properties of the measured system could play would not be restricted to the aforementioned ratio between resolution of \-ex\-pe\-ri\-men\-tal\- construction and quantum threshold, it could comprise also the possibility of defining  dispersion of the measurement outputs 
and it would depend not only on the sensitivity of the \-ex\-pe\-ri\-ment\-, but also on the interaction time (between measuring device and gravitational radiation) 
and on the arm length of the interferometer. In other words, quantum properties of gravitational radiation could play a two-fold role: 
(i) on one hand they could define a quantum threshold, and the ratio between this threshold and the resolution of the apparatus tells us how 
large or small is the backreaction effect of the measuring device upon the wave; (ii) on the other hand, they could also define a dispersion which contains a dependence 
not only on the sensitivity of the experimental construction, but also on $\tau$ and $l$. 

To resume, our results insinuate not only that the current experimental proposals 
could lie very far from their quantum threshold (it is in this sense that these results coincide with the usual hypothesis: 
the backreaction effect of the measuring \-cons\-truc\-tion upon the wave is negligible), but also that these apparatuses could render measurement ouputs that would have a very large dispersion. 
The dependence of the dispersion not only on the sensitivity\- of the device but also on $l$ and $\tau$ can be held 
responsible for the emergence of this feature, and these last two experimental parameters appear in scene because we have considered the quantum properties of the gravitational wave.

It is know [16] that, even if we put aside quantum properties of gravitational \-ra\-dia\-tion\-, the smallness of the intensity of the 
possible sources will make very difficult to extract the information of the gravitational wave out of all the possible noise sources that have the measuring device as origin. The present 
work could then introduce an additional problem to the current proposals, even if we are able to extract the information out of all the noise sources related to the measuring apparatus, 
the \-co\-rres\-pond\-ing\- measurement outputs could have a very large dispersion, i. e. $\delta R <<<1$, and this fact would be a consequence 
of the smallnes of the factor $1/(l^3\tau\Delta I)$. 
Looking at the existing proposals [23, 24], we may see that their experimental parameters would satisfy the 
condition $1/(l^3\tau\Delta I) << 10^5(cm^3s)^{-1}$ (which would not be enough to achieve a small dispersion), and at this point we might wonder if we may neglect the quantum properties of the wave and 
at the same time obtain a realistic picture of the measuring process. 
The present work also suggests that the proposed \-mo\-di\-fi\-ca\-tion\- to these experimental devices [24] (an 
increase in the interaction time between measuring device and gravitational wave) would render not only a larger phase delay in the light field, $\Delta \phi$, but also an increase in the dispersion, namely in $\delta R$.
 
 It has already been pointed out that the concrete form of the weight factor depends on the chosen experimental realization for the measurement process. 
Nevertheless, if the measurement process is realized as a series of repeated short interactions of the measuring device with a subsidiary system 
(meter), the gaussian form of the weight functional will arise automatically [25]. 
Thus, it could be preferable, not only \-ma\-the\-ma\-ti\-cal\-ly\- but also physically, to choose this gaussian as weight factor.

At this point it is noteworthy to mention that concerning the measuring device our approach has been phenomenological, namely the quantum properties 
of the measuring apparatus have not been considered.The complete analysis of the measuring process of gravitational radiation 
must consider the quantum properties of the measuring apparatus and of the gravitational wave. 
The simultaneous introduction of the quantum behaviour of both systems could have as one of its consequences a smaller disparity 
in the order of magnitude between the resolution of the experimental construction and the corresponding 
quantum threshold. We may reformulate this statement and assert that the introduction of the quantum properties of the device could render an increase in the order of magnitude of the quantum threshold, and in consequence a decrease in the dispersion would emerge.

Another question that in this context remains open is the relation of this work with the so called quantum nondemolition 
measurements (QND) [26]. The condition that leads us to QND requires the introduction of a quantum operator, 
which describes the back action of the measuring device upon the measured system, and therefore implies that we must know the quantum behaviour 
of the former [27]. Clearly, our model has, concerning the measuring device, a phenomenological approach and thus does not enable us to 
take into account QND. A possible solution to this problem could be a modification of RPI along Cave's path--integral formalism [28], but this has yet to be done.

\bigskip

\Large{\bf Acknowledgments.}\normalsize
\bigskip

The author would like to thank Deutscher \-A\-ka\-de\-mi\-scher\- Austauschdienst (DAAD) for the fellowship received, M. Mensky for the valuable discussions on the subject, and H. Dehnen and A. Mac{\'\i}as for their support.

\end{document}